# Can current experimental data exclude non-Gaussian genuine band shapes in ultraviolet photoelectron spectra of the hydrated electron?

Ruth Signorell

*Department of Chemistry and Applied Biosciences, Laboratory of Physical Chemistry,*

*ETH Zürich, Vladimir-Prelog-Weg 2, CH-8093, Zürich, Switzerland*

Email: rsignorell@ethz.ch

**Abstract**

Two recent articles present results that allegedly exclude a possible multimodal distribution of the hydrated electron in ultraviolet photoelectron spectra. The first article bases its conclusion on the assumption that the non-Gaussian genuine band shape previously retrieved for the solvated electron in liquid water is an artefact arising from insufficient electron scattering cross sections used in the retrieval. The second article excludes a multimodal band shape based on a photoelectron spectrum of the solvated electron in water clusters recorded at a single ultraviolet photon energy, and it further assumes that cluster results are transferable to the liquid without further justification. Here, we show that based on current data multimodal distributions cannot be unambiguously excluded. Furthermore, the transferability of cluster results to the liquid can neither be justified nor refuted on the basis of currently available experimental ultraviolet photoelectron spectra.

**Liquid phase photoelectron spectra**

A recent article by David Bartels addresses the issue of the band shape of the genuine binding energy (eBE) spectrum of the solvated electron in liquid water [1]. The article states that the genuine eBE of the hydrated electron must be single Gaussian in shape, contrary to the non-Gaussian shape derived in ref. [2] from a simultaneous fit of scattering simulations to eleven experimental liquid photoelectron spectra recorded at eleven different ultraviolet (UV) photon energies below 5.8eV [3]. Ref. [1] concludes that the non-Gaussian ("bimodal") distribution found for the genuine eBE in ref. [2] must have resulted from deficiencies in the scattering cross sections used in the simulations, arising from an allegedly wrong value of $V_0$ = -1.0eV for the escape barrier. Ref. [1] suggests that our scattering cross sections be refitted "with a more realistic choice of $V_0$"; i. e. with $V_0$ closer to ~ -0.1eV; and predicts that the refitted cross sections would result in a genuine eBE spectrum with Gaussian shape.

We have performed the suggested refitting and found that no satisfactory agreement with the set of experimental spectra from [3] could be achieved for $V_0$ = -0.1eV and a Gaussian genuine distribution. More detailed explanations are provided in ref. [4]. Fig. 1 illustrates this for the example of the photoelectron spectrum recorded at 5.8eV photon energy. The black line shows the experimental photoelectron spectrum [2,3] together with a scattering simulation (green line) for an escape barrier of $V_0$ = -1.0eV, the original electron scattering cross sections determined in [2,5,6] and the non-Gaussian genuine band shape derived in [2]. The agreement between experiment and simulation is almost perfect. The red line shows the simulated photoelectron spectrum obtained with cross sections refitted for a barrier $V_0$ = -0.1 eV and a single Gaussian genuine spectrum as suggested in ref. [1]. It is obvious that the simulation with a single Gaussian genuine spectrum and refitted cross sections for a lower escape barrier does not reproduce the experimental spectrum. In ref. [4], we also show that the red spectrum in Fig. 1 is largely identical to a spectrum calculated with the original scattering cross sections [2] and a single Gaussian genuine band shape. This clearly demonstrates that a multimodal distribution is needed to reproduce the high eBE edge of the experimental spectra recorded at higher photon energies (see also figures 3 and S9 in [2]). In other words, the multimodal structure is clearly contained in the experimental spectra, contrary to the statements in ref. [1]. From what we currently now, we can thus exclude deficiencies in the used scattering cross sections as a source for an artificial multimodal distribution. However, as already pointed out in ref. [4], we cannot completely exclude other potential artefacts. Among them could be unknown biases in the experimental UV spectra used in in ref. [2] that might pretend a distorted genuine distribution. Currently, there is not enough evidence to decide beyond any reasonable doubt whether the multimodal genuine distribution in the UV photoelectron spectra of liquid water is a true distribution or caused by artefacts.

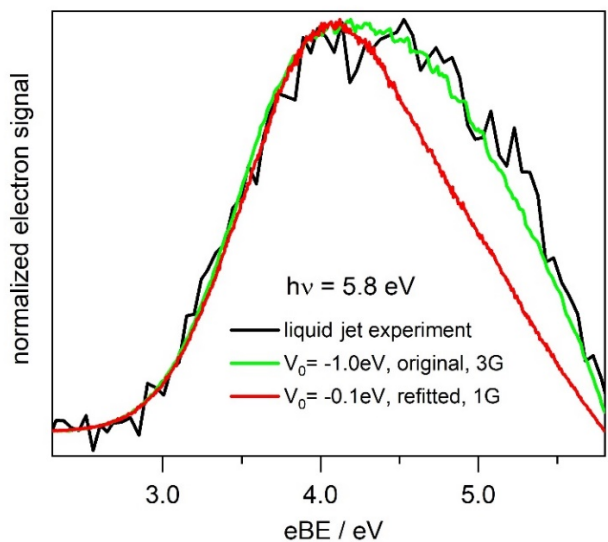

**Fig. 1:** Black line: Experimental eBE spectrum of the solvated electron in liquid water recorded at a photon energy hν = 5.8 eV [3]. Green line: eBE spectrum calculated with the original cross sections, an escape barrier of $V_0$ = -1.0 eV and the multimodal (3G) Gaussian genuine band shape from [2]. Red line: eBE spectrum calculated with cross sections refitted for an escape barrier of $V_0$ = -0.1 eV and a single (1G) Gaussian genuine band shape [4].

**Cluster photoelectron spectra**

Our previous work in ref. [7] had already shown that the photoelectron spectrum of the hydrated electron in large water clusters (with ~300 molecules per cluster) recorded at a single photon energy of 4.66eV is consistent with a single Gaussian genuine band shape. In the same work, we have also shown that this 4.66eV photoelectron spectrum does not allow us to exclude a multimodal ("bimodal") genuine band shape, i. e. a non-Gaussian band shape [7,8]. This is again visualized in Fig. 2, which displays the binned experimental spectrum with error bars (cyan shaded area) together with the cluster simulations for a single Gaussian genuine distribution (black full line) and a multimodal genuine distribution (black dashed line) [7]. Both simulations agree equally well with the experiment. At 4.66eV, photons simply do not have enough energy to cover the non-Gaussian part of the genuine eBE spectrum derived in ref. [2]. These photons thus essentially probe only the dominant Gaussian component of the genuine spectrum (see for clusters Fig. 2, figure 2 in [7] and for the liquid figures 3, S5, and S9 in [2]). This is in contrast to the photoelectron spectrum recorded at 5.8eV, which probes the entire multimodal genuine spectrum (Fig. 1). Fig. 2 also compares our cluster spectrum with corresponding liquid jet spectra from refs. [2, 3] recorded at a photon energy around 4.6eV. The cluster spectrum coincides with the 4.6eV liquid spectrum within the error, while the 4.8eV liquid spectrum seems to deviate slightly from the cluster experiment on the high eBE edge. It would be rash to take such similarities as evidence for the transferability of cluster results to the liquid. In general, differences between clusters and liquid are not implausible, e.g. if one considers that cluster temperatures in supersonic expansions are lower than those in liquid microjets. They might be hidden in the high eBE part of the distribution cut off at photon energies around 4.6eV.

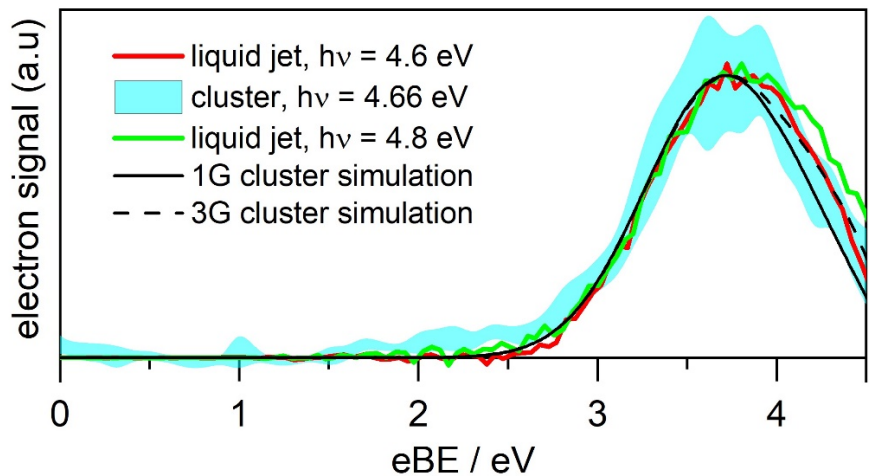

**Fig. 2:** eBE spectra of the hydrated electron at photon energies around 4.6eV. Cyan trace: Experimental cluster spectrum from ref. [7]. The raw data (see figure 2(a) in [7]) were binned with resolution of 0.1eV. The width of the trace indicates three standard deviations from the mean. Red and green trace: experimental liquid spectra from ref. [3] for hv = 4.6eV and 4.8eV, respectively. Black traces: simulated cluster spectra assuming a single (1G) Gaussian genuine band shape (full line) and the multimodal (3G) band shape from ref. [2] (dashed line) as in ref. [7].

In a recent article [9], Svoboda *et al.* claim that they can exclude a multimodal ("bimodal") distribution for the hydrated electron based on a photoelectron spectrum recorded for the solvated electron in water clusters recorded at a single photon energy of 4.66eV (see Fig. 3); i. e. the same energy as used in our cluster work (Fig. 2, [7]). They cite an improved signal to noise ratio of their cluster spectra to support their claim, but do not offer any model predictions of the expected differences between the observable band shapes resulting from single Gaussian and multimodal genuine band shapes. Fig. 2 illustrates that for cluster spectra recorded at 4.66eV the expected difference between a single Gaussian genuine band shape (full line) and the multimodal genuine Gaussian band shape (dashed line) is in fact small, and limited to electron kinetic energies (eKE) below about 0.5eV. Quantitative measurements in this low eKE range are particularly prone to experimental uncertainties. Given the small difference between single Gaussian and multimodal band shapes in a 4.66eV spectrum (Fig. 2), the authors' statement in ref. [9] that their spectra (Fig. 3) "exclude a possible bimodal distribution" appears rather audacious. They also quote liquid results to support their conclusion (see below). However, compared with clusters, the expected difference between a single Gaussian and a multimodal genuine band shape is even less pronounced in the liquid as shown in Fig. 4.

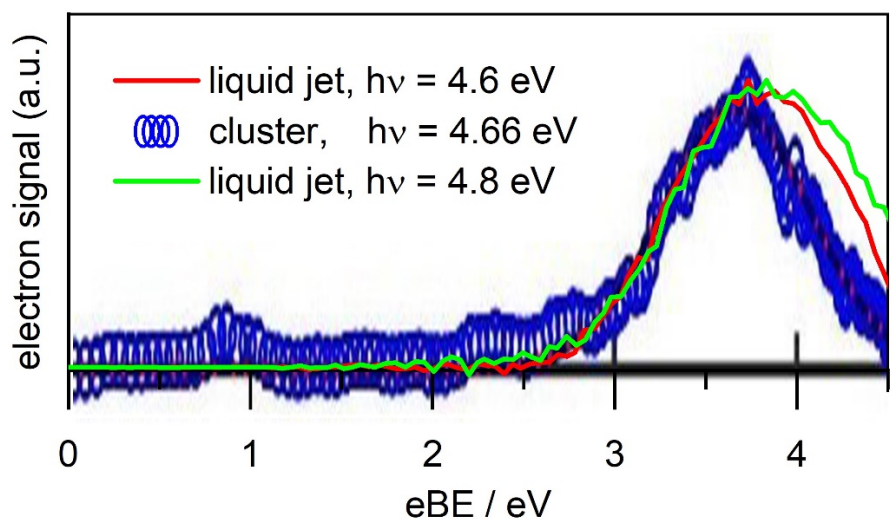

**Fig. 3:** Experimental eBE spectra of the hydrated electron recorded at photon energies around 4.6eV. Blue trace with fit: Cluster spectrum adapted from ref. [9]. The spectrum was copied from figure S4(d) of ref. [9] and adapted under Creative Commons Attribution NonCommercial License

4.0. The liquid spectra are scaled to match the leading (low eBE) edge of the Gaussian fit of ref. [9] (black trace). Red and green trace: liquid spectra from ref. [3] for hν = 4.6eV and 4.8eV, respectively.

The authors of [9] also quote the article by Bartels [1] in support of their claim. As mentioned above, the latter, however, refers to liquid water, not water clusters. It is thus instructive to compare the cluster spectrum of [9] directly with the corresponding liquid jet spectra from refs. [2,3] recorded at photon energies around 4.6eV. The comparison in Fig. 3 shows a marked difference between cluster and liquid in the high eBE (low eKE) range. Even if the exclusion of a possible multimodal distribution of the hydrated electron as stated in [9] were justified for clusters, such a statement would appear not to be simply transferable to the liquid, given the difference between the cluster and the liquid spectrum in Fig. 3. Instead, the conclusion would have to be that the measurements of [9] suggest a significant difference between hydrated electrons in clusters and in the liquid. Therefore, the authors of [9] cannot rely on arguments for the liquid in [1] to support their claim regarding the exclusion of multimodal distributions. On top of that, the arguments put forward in [1] against the multimodal genuine band shape derived in [2] for the liquid have actually proven to be incorrect as discussed here in the context of Figs. 1 and 4, and in ref. [4].

In this context, we note that the justification provided in ref. [1] for a value of the escape barrier around $V_0$ = -0.12eV rather than -1.0eV is actually based on previous sign errors in the derivation, as already pointed out in refs.[4,10,11]. In fact, the cluster data by Coe *et al.* in ref. [12] only provide a lower limit of $V_0 \geq$ -1.72eV; i.e. they favor neither -0.12eV nor -1.0eV. Moreover, it remains unclear whether these cluster data are at all representative for the liquid. A definite value of $V_0$ is still not available for liquid water. We have recently suggested a new droplet experiment that is sensitive to the properties of the escape barrier [13]. This work also highlights that the determination of $V_0$ is more complicated than might appear at first sight. What is important for the current work is that our simulations of the UV spectra of the hydrated electron are largely insensitive to the exact value of $V_0$ as demonstrated in Fig. 1 (see also figure 1 in [4]).

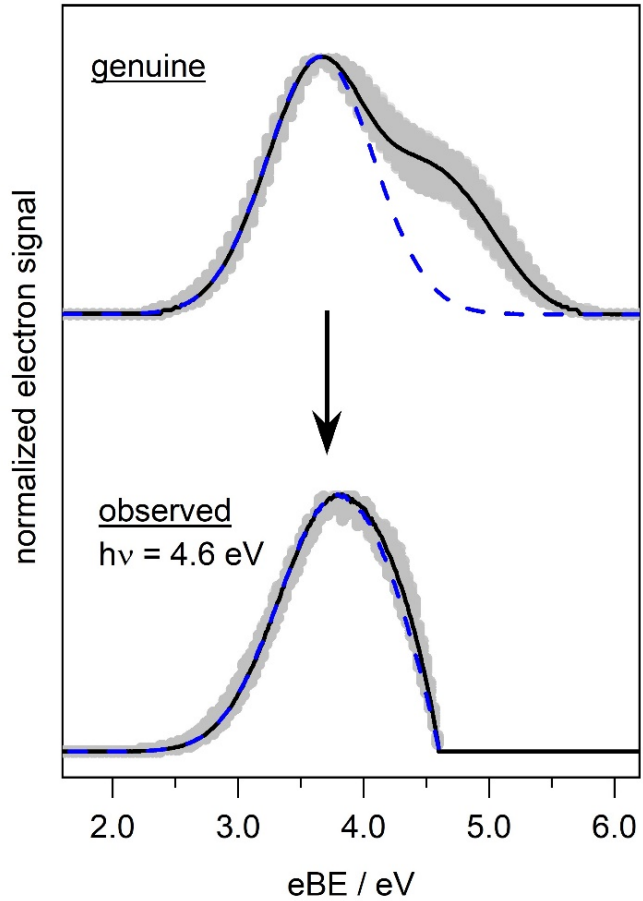

**Fig. 4: Upper spectra**: Genuine eBE spectrum for the hydrated electron in liquid water with 3 Gaussians as derived in ref. [2] (black full line) and 1 Gaussian (blue dashed line). The grey shaded area indicates the fit error determined in [2] for the genuine spectrum. **Lower spectra:** Simulation of observed liquid jet eBE spectra at a photon energy of 4.6eV for the genuine distribution with 3 Gaussians (black full line) and for the genuine distribution with 1 Gaussians (blue dashed line) shown in the upper trace. The simulations employ the original cross sections [2] and an escape barrier of $V_0 = -1.0$ eV. The grey shaded area indicates the uncertainty due to the fit error in the genuine spectrum. The simulated spectra for the two different genuine spectra are almost identical. The figure clearly reveals that spectra observed at 4.6eV cut off the non-Gaussian part of the genuine spectra. This also holds for lower photon energies [2].

**Conclusion**

We do not believe that currently available data for clusters [7,9] recorded at a single photon energy of 4.66eV are sufficient to determine the genuine band shape of the solvated electron in clusters unambiguously - not even to decide whether it is single Gaussian or multimodal. More importantly, the measurements on clusters in [9] do not allow one to judge the band shape in the liquid (Figs. 2, 3 and 4). Generally, the determination of the genuine band shape requires systematic measurements with photon energies covering the whole range well beyond the high eBE edge of the genuine band. While such data is not yet available for clusters, corresponding measurements have already been performed for the liquid up to photon energies of 5.8eV [3]. The resulting liquid jet spectra clearly suggest a non-Gaussian genuine band shape [2]. Contrary to the speculation in [1], this is by no means indicative of deficiencies in scattering cross sections used in the analysis (Fig. 1, [4]). Instead, the deviation from a single Gaussian genuine band shape is directly evident in the experimental spectra of the liquid. Currently available data neither allow one to unequivocally decide whether the genuine distribution of the hydrated electron in the UV is single Gaussian or multimodal, nor do they provide conclusive evidence for the transferability of cluster results to the liquid. For a final resolution of these issues based on experimental data there is no way around further extensive measurements of both clusters and the liquid.